\pdfoutput=1
\documentclass[twocolumn,showpacs,floatfix,prl]{revtex4}

\usepackage[final]{graphicx}
\DeclareGraphicsExtensions{.pdf}
\usepackage{amsmath,amssymb,bm}
\usepackage{float}

\newcommand{\bk}{{\bf k}}

\newcommand{\BGamma}{{\bm \Gamma}}

\begin{document}

\title{Three-dimensional Topological Insulators on the Pyrochlore Lattice}

\author{H.-M. Guo and M. Franz}
\affiliation{Department of Physics and Astronomy, University of
British Columbia,Vancouver, BC, Canada V6T 1Z1}

\begin{abstract}
Electrons hopping on the sites of a three-dimensional pyrochlore
lattice are shown to form topologically non-trivial insulating phases
when the spin-orbit (SO) coupling and lattice distortions are
present. Of 16 possible topological classes 9 are realized for various
parameters in this model. Specifically, at half-filling undistorted
pyrochlore lattice with SO term yields a `pristine' strong topological
insulator with $Z_2$ index (1;000). At quarter filling various strong
and weak topological phases are obtained provided that both SO
coupling and uniaxial lattice distortion are present. Our analysis
suggests that many of the non-magnetic insulating pyrochlores could be
topological insulators.

\end{abstract}

\maketitle

According to recent pioneering theoretical studies \cite{mele1,moore1}
all time-reversal (${\cal T}$) invariant (non-magnetic) band
insulators in three spatial dimensions can be classified into 16
topological classes distinguished by a four-component topological
index $(\nu_0;\nu_1\nu_2\nu_3)$ with $\nu_\alpha=0,1$. Ordinary
`trivial' band insulators have index (0;000) and, in general, possess
no robust surface states. When some of the $\nu$s differ from zero
then the insulator is said to be topologically non-trivial and, as a
result, possesses topologically protected surface states on at least
some of its surfaces. When $\nu_0=1$ surface states exist on {\em all}
surfaces and are in addition robust with respect to weak non-magnetic
disorder. This  is  referred to as a strong topological insulator
(STI). Strong topological insulators are predicted to exhibit a host of
unusual phenomena associated with their non-trivial surface
states. These include proximity-induced exotic superconducting
state with Majorana fermions bound to a vortex \cite{fu1}, spin-charge
separated solitonic excitations \cite{ran1,qi0}, and, in a thin film
geometry, an unconventional excitonic state with fractionally charged
vortices \cite{seradjeh1}. There are also interesting bulk
manifestations of STI physics such the `axion' electromagnetic
response \cite{qi1,essin1} and the topologically protected fermion
modes localized along the core of a crystal dislocation \cite{ran2}.

Topologically nontrivial insulating phases have been predicted
 to occur \cite{bernevig1,fu2,zhang1} and subsequently experimentally
 discovered \cite{konig1,cava1,cava2} in several 2 and 3-dimensional
 crystalline solids. Vigorous  search for new materials in this class is
ongoing.
\begin{figure}
\includegraphics[width=8cm]{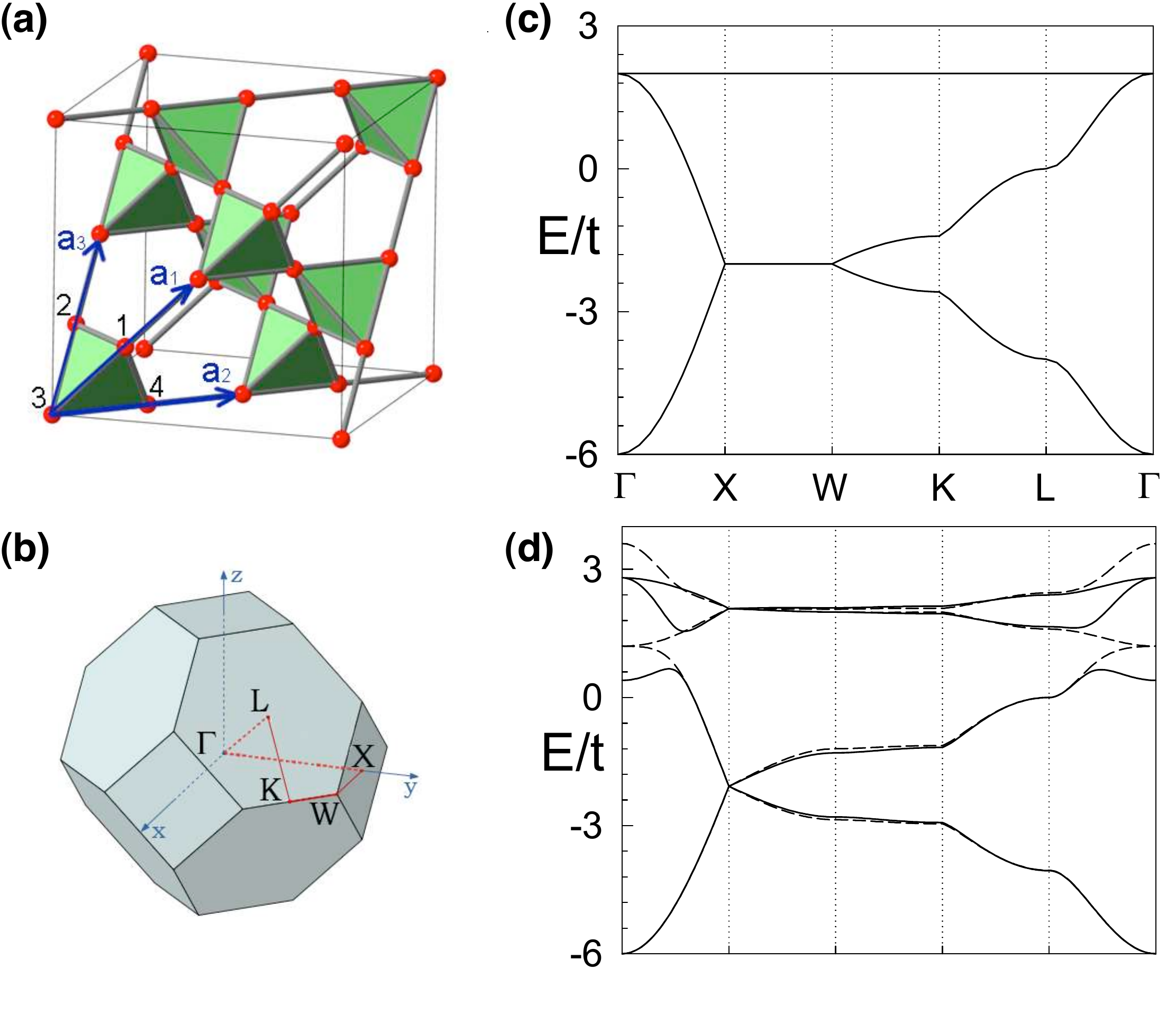}
\caption{(Color online) a) Pyrochlore lattice is a face-centered cubic
Bravais lattice with a 4-point basis forming a shaded tetrahedron.  b)
The first Brillouin zone of the FCC lattice with high symmetry lines
and points indicated. c) Band structure of the tight-binding model
Eq.\ (\ref{h0}). d) Band structure with spin-orbit coupling Eq.\
(\ref{hso}) for $\lambda=-0.1t$ (solid line) and $\lambda=0.1t$
(dashed line). } \label{fig1}
\end{figure}
With the goal of enlarging the space of candidate
crystalline structures that can potentially support topologically
non-trivial insulating phases we study in this Letter a class of
tight-binding models with SO coupling for electrons moving on the {\em
pyrochlore} lattice displayed in Fig.\ \ref{fig1}a. Our model belongs
to the class of 3D `frustrated hopping' models \cite{balents1} and the
motivation for this study comes in part from our recent finding that
electrons on the kagome lattice, a canonical example of the frustrated
structure in 2D, form a 2D topological insulator when SO coupling is
present \cite{guo1}.

 Our main finding here is that, quite generically, whenever electrons
 hopping on the pyrochlore lattice acquire a band gap from SO
 interactions the resulting state is either a STI or a weak
 topological insulator (WTI), defined as a state with $\nu_0=0$ but at
 least one $\nu_{i=1,2,3}\neq 0$. At quarter filling the physics
 leading to the TI behavior on the pyrochlore lattice is somewhat
 similar to the Fu-Kane-Mele (FKM) model on the diamond lattice
 \cite{mele1}. SO interaction produces Dirac-type spectrum at the
 three ${\bf X}$ points of the Brillouin zone (BZ), Fig.\ \ref{fig1}b,
 and uniaxial crystal distortion is needed to open up a gap. The
 resulting Z$_2$ indices are however different from FKM. At half
 filling, the band crossing occurs at the ${\bm \Gamma}$ point and is
 {\em quadratic} rather than Dirac-like. In this case SO coupling by
 itself opens up a gap and no lattice distortion is required. This is,
 to our knowledge, a unique behavior, which produces a highly symmetric
 `pristine' STI characterized by index (1;000).

We now supply the technical details supporting these claims.
Our starting point is the tight-binding model
\begin{equation}\label{h0}
H_0=-t\sum_{\langle ij \rangle\sigma} c^\dagger_{i\sigma}
c_{j\sigma},
\end{equation}
where $c^\dagger_{i\sigma}$ creates an electron with spin $\sigma$
on the site ${\bf r}_i$ of the pyrochlore lattice and $\langle ij
\rangle$ denotes nearest neighbors. In momentum space Eq.\
(\ref{h0}) becomes $H_0=\sum_{{\bf k}\sigma}\Psi^\dagger_{{\bf
k}\sigma}{\cal H}^0_{\bf k}\Psi_{{\bf k}\sigma}$ with $\Psi_{{\bf
k}\sigma}=(c_{1{\bf k}\sigma},c_{2{\bf k}\sigma},c_{3{\bf
k}\sigma},c_{4{\bf k}\sigma})^T$ and ${\cal H}^0_{\bf k}$ of the form
\begin{equation}\label{hk0}
-2t\left(
  \begin{array}{cccc}
    0 & \cos(k_x-k_y) & \cos(k_x+k_z) & \cos(k_y-k_z) \\
    & 0 & \cos(k_y+k_z) & \cos(k_x-k_z) \\
    & & 0 & \cos(k_x+k_y) \\
    & & &0 \\
  \end{array}
\right)\nonumber
\end{equation}
The lower triangle of the matrix is understood to be filled so that
the matrix is hermitian. The spectrum of ${\cal H}^0_{\bf k}$, Fig.\ \ref{fig1}c,
consists of two degenerate flat bands $E_{\bf k}^{(3,4)}=2t$ and two
dispersive bands
\begin{equation}\label{ek0}
E_{\bf k}^{(1,2)}=-2t\left[1\pm\sqrt{1+A_{\bf k}}\right],
\end{equation}
with $A_{\bf
k}=\cos(2k_x)\cos(2k_y)+\cos(2k_x)\cos(2k_z)+\cos(2k_y)\cos(2k_z)$.
$E_{\bf k}^{(2)}$ touches the two flat bands at the ${\bm \Gamma}$ point
and the band crossing is quadratic.
$E_{\bf k}^{(1)}$ and $E_{\bf k}^{(2)}$ touch along the lines located
at the diagonals of the square faces of the BZ.

At the half-filling, bands 1 and 2 are
filled completely, and the two degenerate flat bands are empty. This
state is a gapless band insulator. We now seek terms bi-linear in
the electron operators that lead to the formation of a gap at the
quadratic band crossing point. We focus on
perturbations that do not further break the translational symmetry
of $H_0$ and preserve ${\cal T}$. A natural term to consider is a
SO interaction of the form
\begin{equation}\label{hso}
H_{\rm SO}=i{\lambda}\sum_{\langle\langle ij
\rangle\rangle\alpha\beta}
 ({\bf d}_{ij}^1\times {\bf d}_{ij}^2)\cdot{\bm \sigma}_{\alpha\beta} c^\dagger_{i\alpha} c_{j\beta},
\end{equation}
where $\lambda$ is the SO coupling strength, ${\bf
d}_{ij}^{1,2}$ are nearest neighbor vectors traversed between second
neighbors $i$ and $j$, and ${\bm \sigma}$ is the vector of Pauli
spin matrices. Since ${\bf d}_{ij}^{1,2}$ lie in three-dimensional
space, the Hamiltonian does not decouple for the two spin projections
and in k-space becomes an $8\times 8$ matrix.
Fig.\ \ref{fig1}d shows the spectrum of ${\cal H}^0_{\bf k}+{\cal H}^{\rm
SO}_{\bf k}$.  For $\lambda>0$ it remains gapless but
for $\lambda<0$ a gap $\Delta_{\rm SO}=24|\lambda|$ opens at the ${\bm
  \Gamma}$ point. This peculiar behavior can be understood by studying
the matrix ${\cal H}^0_{\bf k}+{\cal H}^{\rm
SO}_{\bf k}$  at $\bk=0$. It is easy to see that the SO
coupling splits the 6-fold degeneracy  into a two-fold
degenerate level at $2t+16\lambda$ and a four-fold degenerate level at
$2t-8\lambda$. For $\lambda/t<0$ this allows the four-fold degenerate
flat band to split off from the two-fold degenerate dispersive band.
We shall see momentarily that the resulting state is a STI.

Although the SO interaction reduces the degeneracies of bands 1 and 2,
they still touch at three inequivalent Dirac points ${\bf X}_r=2\pi
\hat{r}/a $, where $r=x,y,z$. At quarter-filling band 1 is fully
occupied and it is interesting to ask what ${\cal T}$-invariant
perturbation would open up a gap at the Dirac points. We have been
able to identify two such terms: (i) lattice distortions leading to
anisotropy in the nearest-neighbor hopping amplitudes, and (ii)
modulations in on-site potentials within the unit cell. Both of these
preserve the unit cell, the inversion symmetry and ${\cal T}$.

For the lattice distortions, since there are six hopping amplitudes in
the unit cell Fig.\ \ref{fig1}a, one has many choices. We now describe
four `basic' highly symmetric anisotropy patterns that open up gaps
with equal magnitude at all three Dirac points. We then classify the
resulting insulating phases and argue that this classification is in
fact exhaustive. The basic distortion, labeled by $l=1,2,3,4$, is
obtained by selecting site $l$ in the unit cell and changing
\begin{equation}\label{dist}
t\to t\pm \eta
\end{equation}
The $+$ sign refers to
the six bonds emanating from site $l$ whereas the $-$ sign refers to
all remaining bonds. This can be achieved by deforming the crystal along
the axis passing
through the site $l$ and the center of the tetrahedron.  For pattern
1, the Hamiltonian ${\cal H}^{\rm dis}_{\bf k}$ describing this
modulation takes the form
\[
-2 \eta\left(
  \begin{array}{cccc}
   0 & \cos(k_x-k_y) & \cos(k_x+k_z) & \cos(k_y-k_z) \\
   & 0 & -\cos(k_y+k_z) & -\cos(k_x-k_z) \\
   & & 0 & -\cos(k_x+k_z) \\
   & & & 0 \\
  \end{array}
\right)
\]
for both spin projections. For $l=2,3,4$ the signs in front of the cosine
 terms are permuted in an obvious way.  The full expression for the spectrum of
${\cal H}^0_{\bf k}+{\cal H}^{\rm SO}_{\bf k}+{\cal H}^{\rm
dis}_{\bf k}$ is complicated but it is easy to establish that gaps
$\Delta_{\rm dis}=4|\eta|$ simultaneously open up at all the Dirac
points.

As mentioned above a gap also opens up as a result of on-site
potential modulation. A convenient symmetric choice defines pattern $l$
as $\epsilon_l=3\mu$ and $\epsilon_{k\neq l}=-\mu$ with $\mu$ a
constant.

We now study the topological classes of these insulating phases. As
shown in Ref.\ \cite{fu2} the $Z_2$ topological invariants $(\nu_0;
\nu_1 \nu_2 \nu_3)$ are easy to evaluate when a crystal possesses
inversion symmetry.  The invariants can be determined
from knowledge of the parity eigenvalues $\xi_{2m}({\bm \Gamma}_i)$ of
the 2$m$-th occupied energy band at the 8 ${\cal T}$-invariant momenta
(TRIM) $\BGamma_i$ that satisfy $\BGamma_i = \BGamma_i + {\bf G}$. The 8
TRIM in our system can be expressed in terms of primitive
reciprocal lattice vectors as $\BGamma_{i=(n_1 n_2 n_3)} = (n_1 {\bf
b}_1 + n_2 {\bf b}_2 + n_3 {\bf b}_3)/2$, with $n_j = 0,1$. Then
$\nu_\alpha$ is determined by the product $(-1)^{\nu_0} = \prod_{n_j =
0,1} \delta_{n_1 n_2 n_3},$ and $(-1)^{\nu_{i=1,2,3}} = \prod_{n_{j\ne i}
= 0,1; n_i = 1} \delta_{n_1 n_2 n_3}$, where
$\delta_i=\prod_{m=1}^{N} \xi_{2m}(\BGamma_i)$.

Our model is inversion-symmetric for all perturbations discussed above and so
 we can use this
method to find $\nu$s. If we select site 3 of the unit cell, Fig.\
\ref{fig1}a, as the center of inversion then the parity operator
acts as ${\cal P}[\psi_1({\bf r}),\psi_2({\bf r}),\psi_3({\bf
r}),\psi_4({\bf r})]= [\psi_1(-{\bf r}-{\bf a}_1),\psi_2(-{\bf
r}-{\bf a}_3),\psi_3(-{\bf r}),\psi_4(-{\bf r}-{\bf a}_2)]$ on the
four-component electron wave function in the unit cell labeled by
vector ${\bf r}$. In momentum space and including spin the parity
operator becomes a diagonal $8\times 8$ matrix ${\cal P}_{\bf k}=
{\rm diag}(e^{-i{\bf a}_1\cdot {\bf k}},e^{-i{\bf a}_3\cdot {\bf
k}},1,e^{-i{\bf a}_2\cdot {\bf k}})\otimes {\rm diag}(1,1)$. It is
straightforward to obtain the eigenstates of ${\cal H}_{{
\BGamma}_i}$ and the parity eigenvalues of the occupied bands
numerically, then determine the $Z_2$ invariants.
At half filling,
we find that $\delta=1$ at the $\BGamma$ point and $\delta=-1$ at other
TRIM, so the spin-orbit phase is a $(1;000)$ strong topological
insulator.

At quarter filling, we find 8 Z$_2$ classes, depending on the type of
the distortion.  4 of these are STIs and 4 are WTIs (Table I). The
distinction between STI and WTI is easy to understand on physical
grounds by considering some limiting cases. In the limit
$\eta\to -t$  the electrons can only move
in decoupled parallel planes, each forming a 2D kagome
lattice. Electrons hopping in the kagome lattice with spin-orbit
interaction of the form (\ref{hso}) produce a 2D topological insulator
\cite{guo1}. A collection of such planes results in WTI even after
interplane coupling is restored. When $\eta\to t$, on the other hand,
the resulting structure remains 3-dimensional and STI behavior
prevails.
\begin{table}

\begin{tabular}{|c|c|c|c|c|c|}
  \hline
  Dis & Mass ($m_x, m_y, m_z$)& & $Z_2$ class & & $Z_2$ class \\
  \hline
  1 &$-\eta, \eta, \eta$ & $\eta<0$ & 0;100 & $\eta>0$ & 1;100 \\
  \hline
  2 &$\eta, -\eta, \eta$& $\eta<0$ & 0;001 & $\eta>0$ & 1;001 \\
  \hline
  3 &$\eta, \eta, -\eta$& $\eta<0$ & 0;111 & $\eta>0$ & 1;111\\
  \hline
  4 &$-\eta, -\eta, -\eta$& $\eta<0$ & 0;010 & $\eta>0$ & 1;010\\
  \hline
 \end{tabular}
\caption{$Z_2$ class for the insulators at quarter filling
and the corresponding Dirac mass values in the low-energy effective
 Hamiltonian (\ref{heff}) for different
distortion patterns and arbitrary $\lambda\neq 0$. } \label{z2table}
\end{table}

To develop better understanding for the insulating phases at quarter filling
 we now examine the
form of the low-energy Hamiltonians governing the excitations in
the vicinity of the three Dirac points. This is obtained by
linearizing ${\cal H}^0_{\bf k}+{\cal H}^{\rm SO}_{\bf k}+{\cal
H}^{\rm dis}_{\bf k}$ near ${\bf X}_r$ and subsequently projecting
onto the subspace associated with bands 1 and 2. Near the ${\bf X}_z$ point we
 rescale momenta
as $12\lambda k_x (k_y) \rightarrow k_x(k_y)$ and $4tk_z\rightarrow
k_z$, and obtain a three dimensional Dirac Hamiltonian,
\begin{equation}\label{heff}
{\cal H}_{\rm eff}^z = \tau^x k_z + (\sigma^x k_x + \sigma^y k_y)\tau^y +
2m_{z}^l \tau^z.
\end{equation}
${\cal H}_{\rm eff}^{x,y}$ are the same with $x, y$ and $z$ permuted
in $k_i$ and $\sigma^i$. Index $l$ in the mass labels the distortion
pattern and the values of masses are shown in Table I. We observe that
$l=1,2,3,4$ and two possible signs of $\eta$ exhaust all possible sign
combinations for the three Dirac masses. Since the Z$_2$ index can
only change when at least one of the masses goes through zero
it follows  that our classification in Table I is exhaustive. In
particular given {\em any} cell-periodic pattern of bond distortions
and on-site energies the Z$_2$ class is uniquely determined by the
pattern of the Dirac mass signs listed in Table I.

One can also study the origin of the topologically protected surface
states using the above low-energy Hamiltonian (\ref{heff}).
Consider, for the sake of concreteness, a boundary between two different phases,
running along, say, the $z=0$ plane in real space. We take
distortion pattern 2, $\eta<0$ in the left half-space and
pattern 1, $\eta>0$ in the right half-space. The mass $m_z$
necessarily undergoes a sign change across the $z=0$ boundary. Such
a soliton mass profile is known to produce massless states in the
associated Dirac equation, localized near the boundary \cite{jackiw1}.
Specifically, 3D Dirac equation
\begin{equation}\label{dir1}
[-i\tau^x \partial_z +(\sigma^x
k_x+ \sigma^y k_y)\tau^y  + m(z) \tau^z]\phi_{\bf k}(z) \\
=E\phi_{\bf k}(z)
\end{equation}
with $m(z\to\pm\infty)=\pm m_0$ has gapless solutions
\begin{equation}\label{dir2}
\phi_{\pm\bf k}(z)= \frac{1}{\sqrt{2}}
\left(
               \begin{array}{c}
                 \pm\varphi_\bk \\
                 \pm i\varphi_\bk \\
                 1 \\
                 i \\
               \end{array}
             \right)
e^{-\int_0^zm(z')dz'},
\end{equation}
extended in the $z=0$ plane but localized in the transverse direction,
with linearly dispersing energy $E_{\pm}=\pm k$, where $\varphi_\bk=
(k_x-ik_y)/k$ and $k=\sqrt{k_x^2+k_y^2}$. The number of
gapless states determines the topological class of the phases on two sides of
the boundary. If the number is odd, the boundary is between a WTI
and a STI phase. If the number is even, the boundary is between two
WTI or two  STI phases.

\begin{figure}
\includegraphics[width=8.5cm]{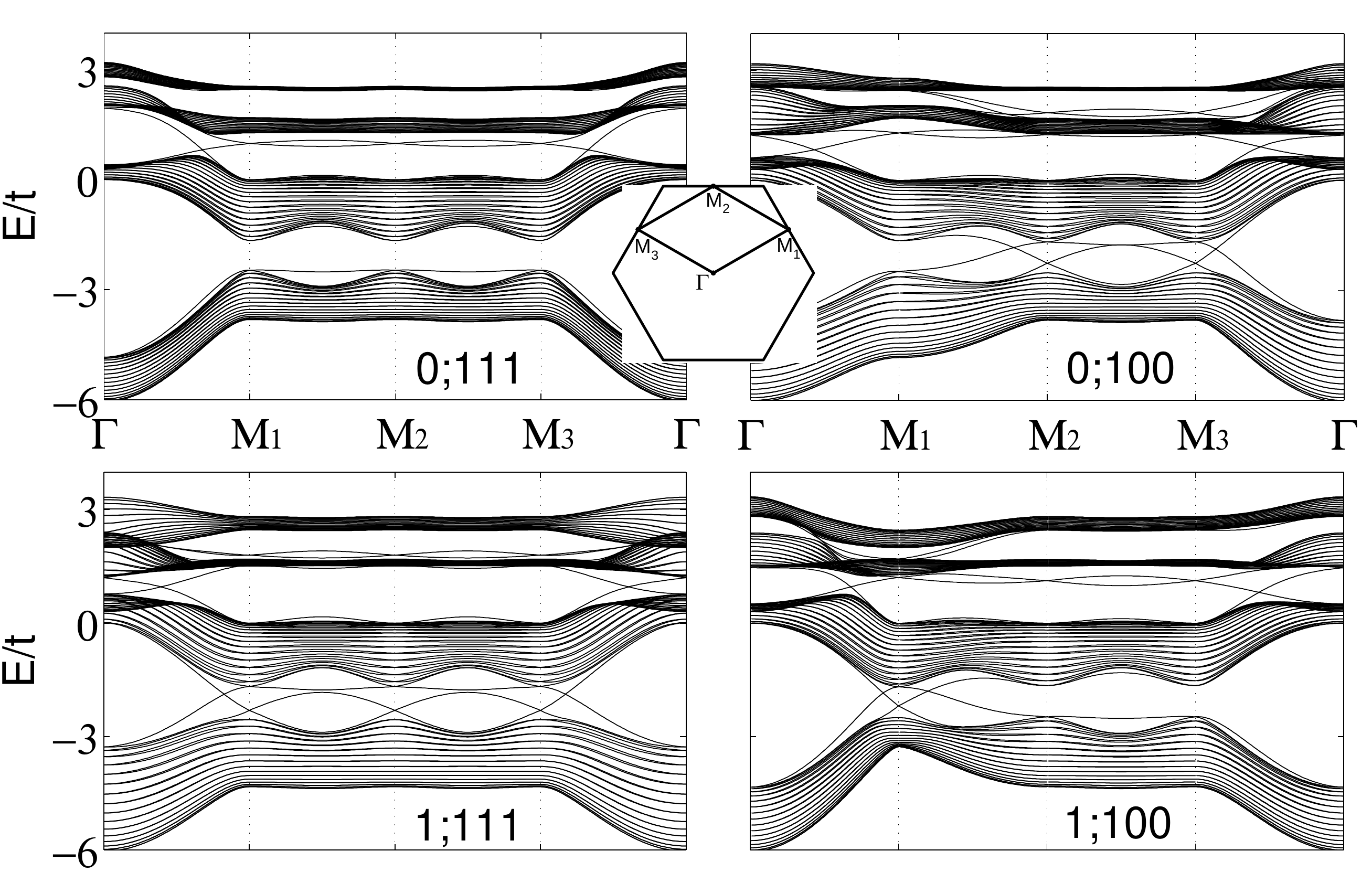}
\caption{Band structures for a slab with a 111 face in various
insulating phases with $\lambda=-0.1t$ and $|\eta|=0.2t$. The Z$_2$
index refers to quarter filling. At half filling all four panels
represent a (1;000) STI. The inset shows the surface BZ with
high-symmetry points marked. } \label{fig2}
\end{figure}
To further support our identification of topological classes given
above, we have performed numerical diagonalizations of the lattice
Hamiltonian $H_0+H_{\rm SO}+H_{\rm dis}$ using the slab
geometry. Fig.\ \ref{fig2} shows the two-dimensional band structures
of a representative sample of the insulating phases obtained for
various patterns of crystal distortions. We consider a slab geometry
with two $111$ surfaces, and plot band energies along lines that
connect the four surface TRIM. Four bulk bands are clearly visible and
there are also surface states some of which traverse the gap.

At half filling the system is in the $(1;000)$ STI phase (we take
$\lambda<0$). Irrespective of the lattice distortion there is a single
Fermi surface (FS) around the $\BGamma$ point for each surface. Since
the two surfaces are inequivalent in this geometry the Fermi surfaces
are also different.

At quarter filling more possibilities arise. In the WTI phases
$(0;111)$ and $(0;100)$ the (even) number of surface FS depends on the
orientation of the surface vector with respect to $(\nu_1\nu_2\nu_3)$,
as discussed in Ref.\ \cite{mele1}. For instance the $111$ surface has
no surface FS in the $(0;111)$ insulator while there are two per
surface in the (0;100) phase.  In the STI phase $(1;100)$ and
$(1;111)$ there are 1 and 3 Dirac points on TRIM, respectively.  Near
each such Dirac point a pair of robust spin-filtered states
exists. Crossings at other momenta can occur, however there is always
an even number of such crossings, confirming the above
arguments.

Many insulating compounds with pyrochlore structure are known to exist
\cite{pyrev}. These follow a formula $A_2B_2$O$_7$ with $A$ typically
a rare earth and $B$ a transition metal element. Existing experimental
studies so far focused mostly on the magnetic pyrochlores due to their
promise as candidate systems for exotic spin liquid and spin ice
ground states brought about by the geometric frustration inherent to
the pyrochlore structure. Our theoretical results show that {\em
  non-magnetic} pyrochlores with strong SO coupling could exhibit
interesting physics. One promising candidate is pyrochlore
Cd$_2$Os$_2$O$_7$ which shows insulating behavior below 225K
\cite{mandrus1}. Band structure calculations in this compound favor
non-magnetic ground state and indicate strong SO effects
\cite{singh1}.  Also promising are the Ir-based pyrochlores
$A_2$Ir$_2$O$_7$ since various Ir-based transition metal oxides have
been predicted and reported to exhibit significant SO effects
\cite{chen1,shitade1,kim1}. In addition, for $A=$Nd, Sm, Eu
metal-insulator transitions have been reported at 36, 117 and 120K,
respectively \cite{matsuhira1}.

Whether a particular pyrochlore is a topological insulator can be
established only through a detailed band structure calculation or an
experimental measurement. These are clearly beyond the scope of our
present study. It is however very encouraging to note that, in a preprint that
appeared after the completion of this work, Pesin and Balents
\cite{pesin1} derived a semi-realistic tight binding model model for
$A_2$Ir$_2$O$_7$ ($A=$Pr,Eu) and found band structures for active orbitals
closely resembling that displayed in Fig. 1d for $\lambda/t <0$. They
find, in agreement with our results, that the system at half
filling is a pristine strong topological insulator.
We conclude that pyrochlore oxides are likely to open a new frontier
in the quest for technologically useful topological insulators and,
more generally, exciting new topological states of quantum
matter. Clearly, detailed band structure calculations and careful
experimental studies of these families of materials are warranted.

\emph{Acknowledgment}.--- Authors are indebted to J. Moore,
G. Rosenberg, B. Seradjeh and C. Weeks for stimulating
discussions. Support for this work came from NSERC, CIfAR and The
China Scholarship Council.


\end{document}